\documentclass[twocolumn,11pt]{article}
\begin{document}
\global\def\refname{{\normalsize \it References:}}
\baselineskip 12.5pt
%
%
%

\title{\LARGE \bf Quantum features of consciousness, computers and brain}

\date{May 12, 2009}

\author{\hspace*{-10pt}
\begin{minipage}[t]{2.7in} \normalsize \baselineskip 12.5pt
\centerline{Michael B. Mensky}
\centerline{P. N. Lebedev Physical Institute, RAS}
\centerline{Leninsky prosp. 53}
\centerline{119991 Moscow}
\centerline{Russia}
\centerline{mensky@lpi.ru}
\end{minipage} \kern 0in
%
\\ \\ \hspace*{-10pt}
\begin{minipage}[b]{6.9in} \normalsize
\baselineskip 12.5pt {\it Abstract:}
Many people believe that mysterious phenomenon of consciousness may be connected with quantum features of our world. The  present author proposed so-called Extended Everett's Concept (EEC) that allowed to explain consciousness and super-consciousness (intuitive knowledge). Brain, according to EEC, is an interface between consciousness and super-consciousness on the one part and body on the other part. Relations between all these components of the human cognitive system are analyzed in the framework of EEC. It is concluded that technical devices improving usage of super-consciousness (intuition) may exist. 
\\ [4mm] {\it Key--Words:}
Consciousness, quantum theory, brain, intuition
\end{minipage}
\vspace{-10pt}}

\maketitle

\thispagestyle{empty} \pagestyle{empty}
%
%
\section{Introduction}
\label{s1} \vspace{-4pt}

One of the mysterious phenomena in the sphere of life is consciousness. It is of course closely connected with thinking and, more generally, with cognitive ability of human. Since thinking is a function of brain, it seems natural to suggest that consciousness is also produced by brain. Many people accept this point of view. This is however not evident because there is an essential difference between the phenomena of thinking and consciousness. Thinking is a direct analogue of computation if the latter is regarded in the broad sense of the word (as including for example logical operations). Consciousness, although well known to everyone, is something that hardly can be clearly defined. 

This is one of the reasons why many attempts have been undertaken to connect consciousness, and more generally, area of psychic, with another mysterious area, quantum mechanics. The latter is a regular branch of science and is therefore well elaborated with respect to its practical applications. However, conceptual basis of quantum phenomena, their radical variation from classical phenomena are not clearly understood up to now. The philosopher David Chalmers formulates the motivation for quantum theories of consciousness (or mind) as follows: ``a Law of Minimization of Mystery: consciousness is mysterious and quantum mechanics is mysterious, so maybe the two mysteries have a common source.'' 

Various ways to connect quantum mechanics with consciousness (or mind) were proposed by Wolfgang Pauli, David Bohm, Roger Penrose, Henry Stapp and other physicists (saying nothing of philosophers). In 2000 the author suggested an approach to this problem based on the Everett's (`many-worlds') interpretation of quantum mechanics. This approach, developed later in a series of papers has been called Extended Everett's Concept (EEC). It seems to be the shortest line of consideration connecting quantum theory with consciousness. 

What makes EEC convincing is that, at the price of only two simple postulates, a great number of mysterious mental phenomena are explained (see Sect.\ref{S2}). The nature of consciousness is not strictly defined in EEC (this is not necessary since the features of consciousness are well defined instead). Yet it is clear that consciousness, or rather complex consisting of explicit consciousness and super-consciousness (manifesting itself in the regime of unconscious), is a human's ability providing the best possible orientation in the world. According to EEC, consciousness is not produced by brain, but is independent of it. The brain serves as an interface between conscious and the body. 

Although consciousness in EEC is directly connected with quantum features of our world, no structure in brain of the type of quantum computer is suggested. Rather the whole quantum world is a sort of quantum computer supporting the phenomenon of consciousness and super-consciousness. Instead of being an origin of consciousness, real quantum computers (even their primitive realizations existing now) can be used to construct models of quantum world demonstrating how the phenomena of life and consciousness may exist. 

Due to special features of human super-consciousness, it cannot be replaced by the action of any technical device or even any material system. However, technical equipment may be used to make usage of super-consciousness more efficient.

\section{Features of consciousness in Extended Everett's Concept (EEC)}
\label{S2} \vspace{-4pt}

It is well known that quantum mechanics suffers from conceptual problems (paradoxes) that are not solved up to date. The reason of these problems is in fact contradiction between linear character of quantum-mechanical evolution and the assumption that during measurement in a quantum system it undergoes to reduction (i.e. the state of the system change so that it correspond to the measurement result). 

This contradiction is absent in the `many-worlds' interpretation of quantum mechanics proposed in 1957 by Everett. This interpretation seems complicated since it is in conflict with our intuition based on classical physics. However, it correctly reflects the quantum concept of reality. It turned out that this interpretation enable one to understand what is our consciousness and explain some strange features of our psychic. The shortest line of consideration leading from quantum theory to theory of consciousness is followed in the Extended Everett's Concept (EEC) proposed by the author in 2000. 

\subsection{Contradiction in quantum mechanics: linearity versus reduction postulate}\vspace{-4pt}

In the generally accept4ed Copenhagen interpretation of quantum mechanics measurement of a quantum system is described by the reduction postulate (suggested by von Neumann). According to this postulate, in the course of measurement the state of the system changes so that it be in accord with the result of the measurement. 

Let for example the  measurement makes distinction of the states $\psi_i$ from each other, and this is done by the measuring device originally in the state $\Phi_0$ and in one of the states $\Phi_i$ after the measurement. This means that the initial state of the measured system and the measuring device $\psi_i\Phi_0$ goes over to $\psi_i\Phi_i$ after the measurement. What then happens if the initial state of the measured system is a superposition $\psi=\sum_i c_i\psi_i$? The initial state of the measured system and measuring device is in this case $\psi\Phi_0=\sum_i c_i\psi_i\Phi_0$. According to the reduction postulate, $i$-th result of measurement will be obtained with the probability $p_i=|c_i|^2$, and the final state of the (system+device) will turn out to be $\psi_i\Phi_i$ (the state of the measuring device corresponds to the $i$-th measurement result). 

This picture of what happens in measurement is very simple and in agreement with our every day experience. Moreover, accepting this postulate, one may be sure that all predictions will be correct (agree with experiment). This is why reduction postulate is accepted by most of physicists. However this postulate and the above simple picture of measurement contradicts to linearity of evolution which is the main feature of quantum mechanics, perfectly confirmed by experiments. 

Indeed, the evolution of the quantum system (measured system + measuring device) during the period of measurement is presented by some evolution operator $U$ (which presents the solution of Schr\"odinger equation). The requirement that the measurement distinct between the states $\psi_i$ may be written as 
$$
U\psi_i\Phi_0=\psi_i\Phi_i
$$
Then directly, from the linear character of the operator $U$, the following evolution law may be derived for the superposition as the initial state: 
$$
U\psi\Phi_0=U\sum_i c_i\psi_i\Phi_0=\sum_i c_i\psi_i\Phi_i
$$
We see that linearity implies that the final state has to be not one of the terms $\psi_i\Phi_i$ (as is assumed by the reduction postulate) but the superposition of all them. 

This is a crucial point. The picture of reduction in measurement turns out to contradict to linearity of evolution in quantum mechanics. This contradiction is an origin of quantum paradoxes, for example known Schr\"odinger cat paradox. 

\subsection{Everett's interpretation}\vspace{-4pt}

Conceptual problems existing in the conventional (Copenhagen) interpretation of quantum mechanics are overcome in Everett's (many-world) interpretation \cite{Everett,DeWittGrah73everett}. The way chosen by Everett is very simple. Instead of accepting reduction postulate (and therefore abandoning linearity in the process of measurement) he assumed that linearity is valid in all processes including measurement. The state of the whole (measured + measuring) system produced in the measurement process (i.e. in the process of interaction of the two subsystems) is taken to be the above superposition $\sum_i c_i\psi_i\Phi_i$. The state of this form is called entangled state (of the measured and measuring systems). 

However, then one discovers that an unexpected feature appears in the Everett's interpretation: typical state of the quantum world is a superposition of classically inconsistent (classically distinct) states, or \textit{classical alternatives}. In the above example the states $\psi_i\Phi_i$ (for various $i$) differ from each other by  the states of the (macroscopic) measuring device. For example, various $i$ may correspond to various positions of the device's pointer. According to our common sense any pair of these classical pictures of the world exclude each other. However, according to the Everett's interpretation they coexist. 

Remark that now, for all further argument, we may forget why we came to this strange conclusion and what is the structure of the state obtained in the course of measurement. The only thing important is that, according to the Everett's interpretation, classically distinct states may coexist in superposition. We shall call such states \textit{classical alternatives}, or \textit{classical projections} of the quantum state of the world. 

It is clear that this strange feature needs justification. It is made agree with the every day experience by the conjecture that \textit{classical alternatives are separated from each other by consciousness}. This means that, while perceiving any of these classical alternatives, the observer does not perceive all the rest as if they were absent. All alternatives are perceived by any observer, but they are perceived separately from each other. 

This may be illustrated by simple formulas. Let us enumerate classical alternatives by index $i$. Then the state of the system (an observer + external world) may be written as the entangled state 
$$
\Psi = \sum_i \Psi_i \chi_i
$$
where $\chi_i$ is the state of the observer perceiving the $i$-th classical alternative, and $\Psi_i$ the state of the external (in respect to the observer) world in the $i$-th classical alternative state. Remark that in this expression the most part of the observer's body may be included in the ``external world'', denoting by $\chi_i$ the state of only its brain (or even of some structure in the brain reflecting the state of the rest of the world). 

Another formulation of the same situation refers to the image of \textit{Everett's worlds} (the term replacing classical alternatives). One may think that the state of the quantum world is adequately presented by the set of classical worlds called Everett's world. the world around an observer is objectively quantum, but  subjectively he perceives it as one of the Everett's classical worlds around him. In each of these worlds just the same observer exists, but the ``copies'' of the same observer know nothing about each other. 

The formulation in terms of the Everett's worlds is considered sometimes more transparent. However, we prefer to speak of the set of classical pictures, or classical projections, of the quantum world. All these projections are perceived by consciousness, but separately from each other. 

\subsection{EEC: the path to theory of consciousness }\vspace{-4pt}

The author put forward the so-called Extended Everett's Concept (EEC)  \cite{MBM2000,MBM2005,MBM2007ufn,mbm2007postcorrect} which allows to go over from the Everett's interpretation of quantum mechanics to some basic points of theory of consciousness. It is accepted in EEC that not only consciousness separate the alternatives but \textit{consciousness is nothing else than the separation of alternatives}. 

This immediately leads to the consequence that the separation of alternatives disappears in the unconscious regime so that one obtains access to all alternatives. Therefore, \textit{in unconscious regime one obtains super-consciousness having access to all classical alternatives}. This not only predicts `supernatural' capabilities of consciousness but also explains why these capabilities reveal themself when (explicit) consciousness is turned off or weakened, for example in dream or meditation (the fact well known in all strong psychological practices). 

This explains not only parapsychology but such well known phenomena as \textit{intuitive guesses} including great \textit{scientific insights}. In fact super-consciousness is a mechanism of \textit{direct vision of truth}. 

The simplicity of derivation of these strange (but many times confirmed) features of consciousness hints that the approach taken in EEC is correct. At the same time this approach does not point out what is the nature of consciousness and super-consciousness so that various philosophical interpretations of them may be accepted. Practically this means that the difference between such philosophical directions as materialism and idealism become relative or completely irrelevant. 

\section{Apparatus has no intuition but may help human to use intuition}
\label{S3} \vspace{-4pt}

According to EEC, conscious (understood broadly, i.e. as an explicit consciousness and super-consciousness) is characteristic feature of life \cite{mbm2007postcorrect}. This makes possible direct (intuitive) vision of truth. This means that a living being can found its actions on the information, part of which is unavailable from the classical picture of the world perceived by it subjectively. This part of information comes from the alternative classical pictures of world included, as components of the superposition, in the whole quantum state of the world. A human may intuitively look for such information in order to make use of it. Primitive living beings exploit such information without being aware of it, but obtaining the corresponding benefit (increasing quality of their life). 

An important question is whether such information (which can be found intuitively) may also be obtained with the help of a sort of technical device (for example classical or even quantum computer). The answer is negative because inanimate material system can have no super-consciousness. 

However, a technical device, or inanimate material system, may be helpful in usage of super-consciousness by human. Let the material system denoted by $\varphi$ is interacting with the human or/and with the external world in such a way that its state entangles with the classical alternatives: 
$$
\Psi=\sum_i \chi_i\varphi_i \Phi_i 
$$
Here $\chi_i$, $\varphi_i$ and $\Phi_i$ are correspondingly the alternative states of the human, of the material system (for example computer) serving as the human's instrument, and of their environment. Various values of the index $i$ correspond to the alternative classical states of the world (various Everett's worlds). 

We see that both the human and its computer have the components corresponding to all values of $i$, i.e. to all alternative classical pictures of the world. Consciously the human may perceive only the components corresponding to a single value of $i$. Subjectively he lives in a certain Everett's world (the world number $i$, so that his state is $\chi_i$). He observes his instrument being in the state $\varphi_i$ and the environment in the state $\Phi_i$. Therefore consciously (subjectively) he cannot obtain information from the alternatives having other numbers, $i'\not = i$. 

In the regime of unconscious, the human may use mechanism of super-consciousness. then he has access to all Everett's worlds (all $i'$, both equal and not equal to $i$). This makes intuitive conclusions available for him. However, 1)~this intuitive conclusions about the environment's states $\Phi_{i'}$ are possible even without the material instrument $\varphi$, and 2)~the instrument itself, without a human, has no super-consciousness and therefore cannot ``transfer'' information from one Everett's world $i'$ to another Everett's world $i$. 

We see finally that, since the phenomenon of super-consciousness cannot exist in technical devices or inanimate material systems, these cannot replace humans in obtaining (intuitive) information from ``other Everett's worlds''. 

One very important remark should be made. Although technical devices do not possess human intuition, they may be helpful for more efficiently usage of human intuition. 

In case of such structure of the world's state the instrument $\varphi$ may help human to take information from ``other classical alternatives''. Indeed, two different situations may exist that can be used in different ways. 1)~If the instrument $\varphi$ and the outer world $\Phi$ interact, then some information about the external world's state $\Phi_i$ is reflected in the state $\varphi_i$ of the instrument. Therefore, exploring, with the help of super-consciousness, the states $\varphi_i$ of the instrument (for various $i$) the human obtains some information about the external world in the alternative states $\Phi_i$. 2)~If the entanglement is caused by interacting the instrument with the human body, then it may be helpful in easier fixing intuitive signals about the external world.

\section{What can quantum computer do?}
\label{S4} \vspace{-4pt}

Quantum computer in the usual sense of this term is an information processing device working in the quantum-coherent regime. For realizing this regime, the set of the degrees of freedom (qubits) included in the information processing should be strictly isolated from its environment. This is the main difficulty for realizing quantum computers (although the requirement of isolation may be weakened by means of the error-correcting codes). 

For readout of the computing results, after the necessary cycle of unitary evolution of the computer, some observables of this quantum system undergo measurement. This causes decoherence of the quantum system and brings the results of computing process into classical form (which may be stored as long as is necessary). 

Unlike classical computer, quantum computer can be used for solving only restricted number of problems, but with much greater speed (because of quantum parallelism, i.e. possibility to parallely process enormous number of data). However, just as a classical computer, quantum computer is inanimate material system and cannot intuitively (super-consciously) acquire information from ``other'' classical alternatives (other Everett's worlds). Direct vision of truth, although based on quantum effects, is feasible only for living beings. 

\section{Quantum computer: model for consciousness}
\label{S5} \vspace{-4pt}

Quantum computer may be used for modeling the `quantum consciousness' as the latter is assumed in EEC. Indeed, according to Everett's interpretation of quantum mechanics, all classical alternatives evolve parallely and independently from each other. It is assumed in EEC (generalizing Everett's interpretation) that `consciousness' is nothing else than this independence (separating the alternatives from each other). The `super-consciousness' is, vice versa, unity of all the alternatives as components of a superposition. Both the separation (independence) of the `alternatives' from each other and their unity in the superposition may be illustrated in a quantum computer as a model. This could experimentally demonstrate at least the fundamental possibility that such `quantum consciousness' may indeed exist (see \cite{MBM2007ufn}). 

This structure may be realized in a quantum computer in the following way. The quantum states evolving in a quantum computer are superpositions with a large number of components. Each superposition component carries some classical information (e.g., a binary number) and the evolution of the entire superposition ensures quantum parallelism, i.e., the simultaneous transformation of all these variants of classical information. In the model of quantum consciousness, individual superposition components can model the alternatives into which the consciousness divides the quantum state. The information contained in each component is a model of an `alternative classical reality', i.e. the alternative state of a living creature and its environment. 

The problem in creating the model of this type is 1)~to formulate a criterion of what will be called survival, and 2)~to select the evolution law such that the evolution of every alternative (superposition component) be predictable, and survival in this evolution be possible (although not guaranteed). 

Of course, the task of constructing such a model is by no means simple, but it is basically solvable using a quantum computer. It is well known that `big' quantum computers, which promise extraordinary new capabilities, have not been realized. However, this applies only to quantum computers with the number of cells of the order of a thousand or more. As for quantum computers with the number of cells around ten, they have already been realized. Evidently, the number of cells attained will increase further, though maybe slowly. It is conceivable that even with these `low-power' quantum computers, which will be constructed in the relatively near future, it will be possible to realize the model of `quantum consciousness'.

\section{Conclusion}
\label{S6} \vspace{-4pt}

We considered in the present paper the approach to theory of consciousness based on the Everett's interpretation of quantum mechanics. The approach called Extended Everett's Concept (EEC) has been proposed in 2000 and elaborated in the subsequent years. The aim of the present paper was to analyze, from the viewpoint of this approach, the role of brain and possibility to replace brain by computer for fulfilling some functions. The main conclusions we came to may be formulated as follows: 

\begin{list}{$\circ$}
{\setlength{\topsep 1pt}
\setlength{\itemsep 1pt}
\setlength{\leftmargin 8pt}
\setlength{\labelwidth 6pt}}
	\item 
Consciousness is the inherent ability of the living beings to perceive alternative classical projections of the objectively quantum world separately from each other. 
	\item 
Super-consciousness, or intuition (existing in the state of meditation, trance  or dream), provides access to all classical alternatives and usage of the obtained information.  
	\item 
Brain, besides solving problems of managing the body, serves also as an interface between consciousness and the body, particularly composing queries for the super-consciousness and interpreting its responses. 
	\item 
Feasible ``artificial intellect'' is a machine for calculations and other intellectual operations, but artificial life (being possible to get information from all classical alternatives) is not feasible by definition. 
	\item 
It is possible to create technical devices which could improve interplay between brain and consciousness and thus increase efficiency of super-consciousness.  
\end{list}

The first concrete considerations on the connection of consciousness with quantum mechanics have been found by Wolfgang Pauli in the course of his collaboration with Karl Yung. In 1952 Pauli wrote in his letter to Rosenfeld (cited according \cite{AtmanPrimas2006} in the translation given by the authors of this paper): ``For the invisible reality, of which we have small pieces of evidence in both quantum physics and the psychology of the unconscious, a symbolic psychophysical unitary language must ultimately be adequate, and this is the far goal which I actually aspire. I am quite confident that the final objective is the same, independent of whether one starts from the psyche (ideas) or from physis (matter). Therefore, I consider the old distinction between materialism and idealism as obsolete.'' It seems that the concept of `quantum consciousness' elaborated in the framework of EEC agrees with these thoughts of Pauli.

\vspace{10pt} \noindent
{\bf Acknowledgements:} \ This work was partially supported by Russian Federation president's grant for leading scientific schools, support \# NSh-438.2008.2. The author acknowleges the fruitful discussion with Wolfgang Baer.



\newpage

\begin{thebibliography}{11}

\vspace{-7pt}\bibitem{Everett} 
H. Everett III, {\em Rev. Mod. Phys.} 29, 454--462, 1957, reprinted in J.~A.~Wheeler and W.~H.~Zurek, editors, \textit{Quantum Theory and Measurement}, Princeton University Press, Princeton, 1983. 

\vspace{-7pt}\bibitem{DeWittGrah73everett}
B. S. DeWitt and N. Graham, editors. \textit{The Many-Worlds Interpretation of Quantum Mechanics}. Princeton University Press, Princeton, 1973.

\vspace{-7pt}\bibitem{MBM2000}
M. B. Mensky, 
Quantum mechanics: New experiments, new applications and new formulations of old questions, 
{\em Physics-Uspekhi} 43, 585-600 (2000). 

\vspace{-7pt}\bibitem{MBM2005} 
M. B. Mensky, Concept of consciousness in the context of quantum mechanics, 
{\em Physics-Uspekhi} 48, 389-409 (2005).  

\vspace{-7pt}\bibitem{MBM2007ufn} 
M. B. Mensky, Quantum measurements, the phenomenon of life, and time arrow: three great problems of physics (in Ginzburg's terminology), {\em Physics-Uspekhi}, 50 (4) 397-407 (2007).

\vspace{-7pt}\bibitem{mbm2007postcorrect}
M. B. Mensky, Postcorrection and mathematical model of life in Extended Everett's Concept, {\em NeuroQuantology}~5, No 4, 363--376 (2007) [http://www.neuroquantology.com]. 

\vspace{-7pt}\bibitem{AtmanPrimas2006}
Harald Atmanspacher and Hans Primas, Pauli's ideas on mind and matter
in the context of contemporary science, 
Journal of Consciousness Studies 13(3), 5-50 (2006)


\end{thebibliography}
\end{document}